\documentstyle[12pt,epsfig,amssymb]{article}
\begin{document}
\setlength{\baselineskip}{0.75cm}
\setlength{\parskip}{0.45cm}
\begin{titlepage}
\begin{flushright}
DO-TH 98/20 \linebreak
September 1998
\end{flushright}
\vskip 0.8in
\begin{center}
{\Large\bf Dynamical QCD Predictions for\\
Ultrahigh Energy Neutrino Cross Sections}
\vskip 0.5in
{\large M.\ Gl\"uck, S.\ Kretzer and E.\ Reya}
\end{center}
\vskip .3in
\begin{center}
{\large Institut f\"{u}r Physik, Universit\"{a}t Dortmund \\
D-44221 Dortmund, Germany }
\end{center}
\vskip 0.5in
{\large{\underline{Abstract}}}

\noindent
Neutrino-nucleon total cross sections for neutrino energies up to ultrahigh
energies (UHE), $E_\nu=10^{12}\ {\rm{GeV}}$, are evaluated within the framework
of the dynamical (radiative) parton model. The expected uncertainties of these
predictions do not exceed the level of about 20 \% at the highest energies
where contributions of parton distributions in the yet unmeasured region around 
$x\simeq 10^{-8}$ to $10^{-9}$ are non-negligible. This is far more accurate 
than estimated uncertainties of about $2^{\pm 1}$ due to ad hoc extrapolations
of parton distributions to $x<10^{-5}$ required for calculating UHE cosmic
neutrino event rates.  
\end{titlepage}
%
\noindent
Calculations of ultrahigh-energy (UHE) neutrino-nucleon total cross sections,
relevant for neutrino astronomy, have been improved \cite{1,2,3,4} by taking
into account new high energy measurements of deep inelastic lepton-nucleon
scattering (DIS) at DESY-HERA \cite{5}. These updated calculations were even
further improved very recently \cite{6} not only due to additional small-$x$
precision measurements at HERA \cite{7} which became available in the meantime,
but mainly due to new ideas about the flux of neutrinos \cite{8} from active
galactic nuclei (AGNs) \cite{9}, gamma ray bursts (GRBs) \cite{10} or from
decays of exotic heavy particles of generic top-down (TD) or topological
defects models \cite{11}. Most of these new developments refer to neutrino
energies above about $10^8\ {\rm{GeV}}$ where (anti)neutrino-nucleon cross
sections become sensitive to parton densities at ultrasmall values of
Bjorken-$x$, $x<10^{-5}$, not accessible by present DIS experiments. Different
{\it{assumptions}} about the $x\rightarrow 0$ behavior then lead to different
cross sections. Thus a significant uncertainty in estimating the detectability
of extraterrestial UHE neutrinos with present and future neutrino telescopes
is due to the small-$x$ extrapolations and the resulting uncertainty reaches
typically a factor $2^{\pm 1}$ around $10^{20}\ {\rm{eV}}$ \cite{2,6}. 

The highest presently available $ep$ energy at HERA, $\sqrt{s_{ep}}=314\
{\rm{GeV}}$, is rather small as compared to the envisaged UHE neutrino nucleon
collision energies of up to about $\sqrt{s_{\nu N}}=10^6\ {\rm{GeV}}$. Estimates
of the corresponding total $\stackrel{(-)}{\nu} N$ 
cross sections thus afford either
extensive, possibly unreliable, extrapolations \cite{1,2,3,4,6} of existing
data ($x\gtrsim 10^{-5}$) and their respective fits, or the application 
\cite{3} of QCD inspired models which proved to provide reliable high energy
predictions \cite{12,13,14} in the past \cite{5,7}. In the present work we
shall adopt this second option and base our predictions on calculations within
the framework of the radiative parton model \cite{12,13,14,15} which allows to
calculate the small-$x$ ($x\lesssim 10^{-2}$) behavior of parton densities
from first principles, i.e.\ QCD dynamics, independently of any free (fit)
parameters in the small-$x$ region. These unique dynamical predictions result
from valence-like gluon and sea input densities $xg(x,Q_0^2)\sim x^a$,
$x{\bar{q}}(x,Q_0^2)\sim x^{a^\prime}$ with $a,a^\prime >0$ as $x\rightarrow 0$
at some low momentum scale $Q_0\simeq0.5-0.6\ {\rm{GeV}}$. The resulting
small-$x$ behavior is furthermore perturbatively stable and 
unique at the relevant momentum 
scales $Q^2\simeq M_W^2$. Here we shall apply the parton
distributions in \cite{14}, and in particular the most recent ones in \cite{15}
which take also into account all recent high precision measurements at HERA
\cite{7,16}, to evaluate 
$\sigma^{\stackrel{(-)}{\nu}N} (E_{\stackrel{(-)}{\nu}} )$ 
for $10^2 \lesssim E_{\stackrel{(-)}{\nu}}
\lesssim 10^{12}\ {\rm{GeV}}$. At highest neutrino energies, this requires the
knowledge of parton densities down to $x\simeq 10^{-8}$ ($x \gtrsim
M_W^2/2ME_\nu$ due to the $W$-propagator, with $M$ being the nucleon mass) at
scales $Q^2\simeq M_W^2$. In ref. \cite{15} we have explicitly demonstrated
that our next-to-leading order (NLO) results in this extremely small-$x$ region
are perturbatively stable when compared with the leading order (LO) ones. It
should be kept in mind that this stability refers always to measurable
quantities like structure functions or cross sections rather than to the
auxiliary, not directly measurable, NLO parton distributions \cite{14,15}
since perturbative stability usually requires $f^{NLO}(x,Q^2)\neq
f^{LO}(x,Q^2)$ for $f=q,{\bar{q}},g$. Furthermore, the NLO predictions are, in
contrast to the LO ones, obviously rather insensitive to the specific choice
for the renormalization scale $\mu_R$ appearing in $\alpha_s(\mu_R^2)$ and for
the factorization scale $\mu_F$ appearing in the parton densities
$f(x,\mu_F^2)$ \cite{15}. We shall therefore mainly concentrate on NLO analyses
for predicting the various UHE total cross sections relevant for neutrino
astronomy. 

The above results lend sufficient confidence in the reliability of the
perturbatively calculated UHE $\stackrel{(-)}{\nu} N$ 
total cross sections, $N=(p+n)/2$,
evaluated according to
\begin{equation}
\sigma^{\stackrel{(-)}{\nu}N}(s)=\int_0^1 dx\ \int_0^1 dy\
d^2\sigma^{\stackrel{(-)}{\nu}N}/dx dy
\end{equation}
with
\begin{equation}
\frac{d^2\sigma^{\stackrel{(-)}{\nu}N}}{dx dy} = \frac{G_F^2 s}{2 \pi}\
(1+xys/M_W^2)^{-2} \big[ (1-y) F_2^{\stackrel{(-)}{\nu}} + y^2 x
F_1^ {\stackrel{(-)}{\nu}}\pm y(1-\frac{y}{2}) x F_3^{\stackrel{(-)}{\nu}} \big]
\end{equation}
where $F_i=F_i(x,Q^2=xys)$, $s=2ME_{\stackrel{(-)}{\nu}}$ and $G_F=1.1663\times
10^{-5}\ {\rm{GeV}}^{-2}$. It should be noted that the results for the total
cross sections are insensitive to the lower integration limits in (1) since
the $W$-propagator in (2) restricts $Q^2=xys$ to values around $M_W^2$. Anyway
, as soon as the integrations correspond to scales $Q^2<Q_0^2$, we freeze the
respective parton densities at their given input scale $Q^2=Q_0^2$. 
The $\stackrel{(-)}{\nu} N$ structure functions $F_i$ in (2) can be decomposed as
\begin{equation}
F_i=F_i^{light}+F_i^h
\end{equation}
where $F_i^{light}$ refers to all light ($u,d,s$) contributions and $F_i^h$
denotes all relevant contributions related to heavy quark ($c,b,t$)
transitions. 

For charged current (CC) $\nu_l N\rightarrow l^- X$ reactions we have in
LO-QCD
\begin{eqnarray} \nonumber
F_1^{\nu,light} &=& \frac{1}{2} ({\bar{u}}+{\bar{d}}) + \frac{1}{2}
(d+u)|V_{ud}|^2 + s|V_{us}|^2 \\ 
F_2^{\nu,light} &=& 2 x F_1^{\nu,light} \\ \nonumber
F_3^{\nu,light} &=& - ({\bar{u}}+{\bar{d}})+(d+u)|V_{ud}|^2 + 2s|V_{us}|^2
\end{eqnarray}
where $u=u(x,Q^2)$ etc., except stated otherwise, $s={\bar{s}}$
and the CKM mixing matrix elements are taken from \cite{17}. The heavy flavor
component $F_i^h$ in LO consists of the charm contribution $F_i^c$ according
to the $W^+ s\rightarrow c$ and $W^+ d\rightarrow c$ transitions as described, 
for example, in \cite{18}, and of the top-bottom contribution
$F_i^{t{\bar{b}}}$ according to the $W^+ g\rightarrow t{\bar{b}}$ fusion
subprocess \cite{19}, using a factorization and renormalization
scale equal to $(m_t+m_b)$
with $m_t=175\ {\rm{GeV}}$ and $m_b=4.5\ {\rm{GeV}}$. [Note that the sign
convention for $F_3$ in \cite{19} is opposite to the one in eq.\ (2)]. We
shall compare these heavy quark contributions also with the less adequate, but
frequently used (e.g., \cite{1,2,3,4,6}) massless approach (the so called
'variable flavor' scheme) where intrinsic 'heavy' quark densities are purely
radiatively generated using the ordinary massless evolution equations,
starting at $Q=m_h$. In this case, $F_i^h(x,Q^2)$ in (3) is given by
\begin{eqnarray}\nonumber
F_1^{\nu,h}&=&\frac{1}{2}(d+u)|V_{cd}|^2+s|V_{cs}|^2+c + 
b(\xi,Q^2+m_t^2)|V_{tb}|^2\\
F_2^{\nu,h}&=&x(d+u)|V_{cd}|^2+2 x s|V_{cs}|^2+ 2 x c + 
2 \xi b(\xi,Q^2+m_t^2)|V_{tb}|^2\\ \nonumber
F_3^{\nu,h}&=&(d+u)|V_{cd}|^2+2 s|V_{cs}|^2- 2 c + 
2 b(\xi,Q^2+m_t^2)|V_{tb}|^2
\end{eqnarray}
with $\xi=x(1+m_t^2/Q^2)$
and where we neglected $m_c^2/Q^2$ contributions as
well as $t(x,Q^2)$. We choose the scale $Q^2+m_t^2$ in $b$, since it resulted
in the best perturbative LO/NLO stability in the $c{\bar{s}}$ sector
\cite{18}. For this latter analysis in the so called 'variable flavor' scheme
we shall use the dynamical parton densities of ref.\ \cite{13}. The relevant
structure functions for CC ${\bar{\nu}} N$ reactions in (2) are simply
obtained from the above expressions via $F_{1,2}^{\bar{\nu}}=F_{1,2}^{\nu}(q
\leftrightarrow {\bar{q}})$ and $F_{3}^{\bar{\nu}}=-F_{3}^{\nu}(q 
\leftrightarrow {\bar{q}})$. 

In NLO-QCD all above LO parton densities have to be replaced by the NLO ones
\cite{14,15}, with the appropriate addition of convolutions with the fermionic
Wilson coefficient $C_q$ and the NLO contribution $C_g \otimes g$, as
summarized for example in \cite{20}. It turns out that these additional
convolutions contribute less than about 2\% and thus these complications can
be safely neglected for NLO calculations of high-energy total 
$\stackrel{(-)}{\nu} N$ cross
sections. The heavy flavor piece $F_i^h$ receives additional massive
${\cal{O}}(\alpha_s)$ contributions according to the $W^+ g \rightarrow c
{\bar{s}}$, $W^+ s^\prime\rightarrow g c$, etc., subprocesses \cite{18,21,22}
which we include according to the (${\overline{\rm{MS}}}$) results of  
\cite{18}. The
fact that the $t{\bar{b}}$ production has so far only been calculated in LO
($W^+ g\rightarrow  t{\bar{b}}$) is of minor importance, since the
$c{\bar{s}}$ sector dominates over the much heavier $t{\bar{b}}$ one. It
should be noted that this more appropriate way of calculating the heavy quark
contributions in fixed order perturbation theory without neglecting mass
effects, usually referred to as 'fixed flavor' factorization scheme (also
adopted in \cite{14,15}), results in perturbatively stable LO/NLO
QCD-predictions \cite{18} even for the highest energies \cite{23} attainable
with present cosmic ray shower experiments ($\sqrt{s}=10^4-10^5\ {\rm{GeV}}$).

For completeness we also summarize briefly the structure functions in (2) for
neutral current (NC) reactions, where obviously $M_W\rightarrow M_Z$.  
For $\nu N\rightarrow \nu X$ the light quark
contributions in LO are
\begin{eqnarray} \nonumber
2 F_1^{\nu,light} &=& \frac{1}{2} (u+{\bar{u}}+d+{\bar{d}}) (V_u^2+A_u^2)+
\frac{1}{2} (u+{\bar{u}}+d+{\bar{d}}+ 4 s) (V_d^2+A_d^2)\\
F_2^{\nu,light} &=& 2 x F_1^{\nu,light}\\ \nonumber
2 F_3^{\nu,light} &=& 2(u_v+d_v)(V_u A_u+V_d A_d)
\end{eqnarray}
where $q_v=q_v(x,Q^2)\equiv q-{\bar{q}}$ and $V_u=\frac{1}{2}-\frac{4}{3}
\sin^2\theta_W$, $V_d=-\frac{1}{2}+\frac{2}{3}\sin^2\theta_W$,
$A_u=-A_d=\frac{1}{2}$ with $\sin^2\theta_W=0.232$. The (massive) heavy flavor
contributions to $F_{1,2}^{\nu,h}$ derive from the subprocesses $Z^0
g\rightarrow h {\bar{h}}$, $h=c,b$ and are given in \cite{19}, with
$F_3^{\nu,h}=0$. (The small contribution from $t{\bar{t}}$ production will be
neglected). If heavy quarks are treated less adequately as massless intrinsic
partons, we have instead of eq.\ (5),
\begin{eqnarray} \nonumber
&2 F_1^{\nu,h}&=\ 2 c\ (V_u^2+A_u^2)+2 b\ (V_d^2 +A_d^2) \\
&F_2^{\nu,h}&=\ 2x\ F_1^{\nu,h}\ \  ,\ \  F_3^{\nu,h}=0\ \ .
\end{eqnarray}
In this massless case, $c=c(x,Q^2)$ etc., we use again the 'variable flavor'
parton densities of ref.\ \cite{13}. For NC ${\bar{\nu}} N \rightarrow
{\bar{\nu}} X$ reactions, one simply has $F_i^{\bar{\nu}}=F_i^{\nu}$. In NLO
all above LO parton distributions have to be replaced by the NLO ones
\cite{14,15}, with the additional convolutions $C_q \otimes q$ and $C_g
\otimes g$ \cite{20} being again negligible as for CC reactions.\footnote{
Since the fully massive NLO contributions to $Z^0 g \rightarrow h {\bar{h}}$
have not yet been calculated, we use the LO($\alpha_s$) expressions for the NLO
analysis as well. (The factorization scale is taken to be $\mu^2_F=4m_h^2$
which provides the best perturbative LO/NLO stability for photon-induced
processes $\gamma^\ast g \rightarrow h {\bar{h}}$, etc.\ \cite{23} -- a
stability which holds down to $x=10^{-8}$). The same holds for the above CC
contribution due to $W^+ g \rightarrow t {\bar{b}}$. This approximation is of
minor importance, since the main contributions come, in both CC and NC cases,
from the perturbatively stable light $u,d,s$ quarks and the $c {\bar{s}}$
sector.}

Our relevant most recent NLO(${\overline{\rm{MS}}}$) sea and gluon densities
\cite{15} at $Q^2=M_W^2$ are shown in fig.\ 1, where they are also compared
with our previous GRV 94 \cite{14} predictions\footnote{
The simple analytic parametrizations of the LO and NLO QCD-evolution GRV 94
predictions for parton densities presented in \cite{14} are sufficiently
accurate down to the lowest relevant $x\simeq 10^{-8}$. Moreover, the NLO(DIS)
results are practically the same as the NLO (${\overline{\rm{MS}}}$) ones for GRV 94
\cite{14} as well as for the more recent GRV 98 \cite{15} distributions.}
which are slightly steeper for $x<10^{-5}$, relevant for $\nu N$ cross
sections at $E_\nu> 10^8\ {\rm{GeV}}$. For illustration we also show the input
of the valence-like NLO GRV 98 densities at the input scale $Q_0^2 = 0.40\
{\rm{GeV}}^2$ which becomes vanishingly small at $x<10^{-2}$, particularly for
the gluon input. This illustrates the purely dynamical (i.e.\ parameter-free)
origin of the very small-$x$ structure of sea quark densities and in
particular of the gluon density at $Q^2>Q_0^2$ which dominates the $Q^2$
evolution of ${\bar{q}}(x,Q^2)$ in the small-$x$ region. Also noteworthy is
the stability of ${\bar{u}}+{\bar{d}}$ at $Q^2=M_W^2\gg Q_0^2$ with respest to
our previous GRV 94 results. For comparison the expectations based on the
CTEQ3-DIS \cite{24} and CTEQ4-DIS \cite{25} parton densities\footnote{
These densities have been extrapolated to low-$x$ ($x\lesssim 10^{-5}$) beyond
the validity of the fitted parametrizations ($x>10^{-5}$) using some
reasonable analytic functions. As in the case of the CTEQ3 densities
\cite{24}, the CTEQ collaboration provided us with similar extrapolation
functions for the CTEQ4 distributions \cite{25}. We thank Wu-Ki Tung for a
helpful correspondence.}
are shown as well. Whereas the CTEQ3 extrapolations seem to overestimate the
very small-$x$ region, $x<10^{-5}$, the CTEQ4 ones are in reasonable agreement
with our dynamical expectations.

The resulting CC 
total $\nu N$ cross sections are presented in fig.\ 2 which demonstrates the
stability of the QCD predicted cross sections in the UHE region. The heavy
quark contributions are also shown separately: The $c s^\prime$ sector
contributes significantly, whereas the much heavier $t {\bar{b}}$ contribution
is small. These exactly calculated massive results are also compared with the
results obtained from the less adequate description of treating heavy quarks
as massless intrinsic partons according to eq.\ (5), using the GRV 92
densities \cite{13}. Although this latter simplified massless approach to
'heavy' quark effects {\it{over}}estimates their contributions, their
difference with respect to the exact massive treatment becomes rather marginal
at scales as large as $Q^2\simeq M_W^2$. It should be mentioned that in
practically all conventional approaches where parton densities are obtained
from fitting appropriate parametrizations to presently available DIS data, to
be discussed below, heavy quarks are treated as massless intrinsic
partons. This is also the case for the GQRS 98 results \cite{6} shown in fig.\
2 which are based on the CTEQ4-DIS parton distributions as the nominal
distributions for calculating neutrino-nucleon cross sections and estimating
astrophysical UHE neutrino event rates. It should be kept in mind that these
results have been obtained by using some ad hoc 'brute force' extrapolation to
$x<10^{-5}$, i.e.\ by extrapolating the CTEQ4-DIS densities below $x=10^{-5}$
with the {\it{same}} power in $x$ predicted at $x=10^{-5}$ and
$Q^2=M_W^2$ \cite{6}\footnote{
We thank Ina Sarcevic for a clarifying correspondence on this point.}
. This overestimates the light sea quark densities at $x\simeq 10^{-8}$ by
almost 20\% as compared to the CTEQ extrapolations \cite{25}\footnotemark[3] 
shown
in fig.\ 1 and thus also overestimates the cross section by the same amount at
$E_\nu\simeq 10^{12}\ {\rm{GeV}}$. Nevertheless, the expected total cross
sections agree, somewhat accidentally, rather well with our dynamical
predictions and are thus perfectly legitimate for practical applications. 

To test the amount of extrapolation in $x$ and $Q^2$ involved in these
calculations of total cross sections, fig.\ 3 shows the contribution to
$\sigma_{CC}^{\nu N}$ from different regions of $x$, i.e.\ by choosing a
finite lower limit of $x$-integration $x\geq x_{min}$ in eq.\ (1): Even at
highest neutrino energies the $W$-propagator in (2) constrains \cite{1,2} the
relevant values of $x$ to be larger than about $10^{-9}$. For similar reasons
the relevant $Q^2$ is restricted to values near $M_W^2$ for $E_\nu \gtrsim
1\ {\rm{TeV}}$. Since our dynamical QCD small-$x$ predictions for $x\lesssim
10^{-2}$ agree with all present DESY-HERA measurements down to $x\simeq
10^{-5}$, as discussed at the beginning, our dynamical predictions for UHE
$\nu({\bar{\nu}})$ nucleon cross sections, which are dominated by $x\gtrsim
10^{-8} - 10^{-9}$, appear to be reasonably reliable. 

Similar results hold for $CC$ ${\bar{\nu}} N$ and for the NC
$\nu({\bar{\nu}})N$ total cross sections. Our final 
dynamical NLO high energy (small-$x$)
predictions, based on our
GRV 98 parton distributions \cite{15}, are given within 5 to 10\% by
\begin{eqnarray} \nonumber
\sigma_{CC}^{\nu N} &=& \left\{ 
\begin{array}{l} 1.10\times 10^{-36} {\rm{cm}}^2 (E_\nu/1{\rm{GeV}})^{0.454}\ \ ,
10^5\lesssim E_\nu \lesssim 10^8 {\rm{GeV}} \\
5.20\times 10^{-36} {\rm{cm}}^2 (E_\nu/1{\rm{GeV}})^{0.372}\ \ ,
10^8\lesssim E_\nu \lesssim 10^{12} {\rm{GeV}}
\end{array}\right. \\ \nonumber
\sigma_{CC}^{\bar{\nu} N} &=& \left\{ 
\begin{array}{l} 6.65\times 
10^{-37} {\rm{cm}}^2 (E_{\bar{\nu}}/1{\rm{GeV}})^{0.484}\ \ ,
10^5\lesssim E_{\bar{\nu}} \lesssim 10^8 {\rm{GeV}} \\
5.20\times 10^{-36} {\rm{cm}}^2 (E_{\bar{\nu}}/1{\rm{GeV}})^{0.372}\ \ ,
10^8\lesssim E_{\bar{\nu}} \lesssim 10^{12} {\rm{GeV}}
\end{array}\right. \\ \nonumber
\sigma_{NC}^{\nu N} &=& \left\{ 
\begin{array}{l} 3.55\times 10^{-37} {\rm{cm}}^2 (E_\nu/1{\rm{GeV}})^{0.467}\ \ ,
10^5\lesssim E_\nu \lesssim 10^8 {\rm{GeV}} \\
3.14\times 10^{-36} {\rm{cm}}^2 (E_\nu/1{\rm{GeV}})^{0.349}\ \ ,
10^8\lesssim E_\nu \lesssim 10^{12} {\rm{GeV}}
\end{array}\right. \\ \nonumber
\sigma_{NC}^{\bar{\nu} N} &=& \left\{ 
\begin{array}{l} 3.04\times 10^{-37} {\rm{cm}}^2 (E_{\bar{\nu}}/1{\rm{GeV}})^{0.474}\ \ ,
10^5\lesssim E_{\bar{\nu}} \lesssim 10^8 {\rm{GeV}} \\
3.14\times 10^{-36} {\rm{cm}}^2 (E_{\bar{\nu}}/1{\rm{GeV}})^{0.349}\ \ ,
10^8\lesssim E_{\bar{\nu}} \lesssim 10^{12} {\rm{GeV}}\ \ \ .
\end{array}\right. 
\end{eqnarray}

In fig.\ 4 we compare our most recent predictions for $\sigma_{CC}^{\nu N}$ as
a representative example, based on the GRV 98 distributions, with calculations
based on different sets of parton densities and small-$x$ extrapolations. It
should be again pointed out that all conventionally fitted sets of parton
densities are extrapolated to low-$x$ ($x<10^{-5}$) beyond the validity of the
fitted parametrizations at $x\gtrsim 10^{-5}$ using either some 'reasonable'
analytic functions\footnotemark[3] 
\cite{24,25} or, in most cases, a fixed slope of
$x{\bar{q}}(x,M_W^2)$ at the lowest $x$ permitted by present HERA experiments
($x\simeq 10^{-5}$). As already demonstrated in fig.\ 1, the dynamical GRV
94 densities result in very similar high energy predictions as the GRV 98
ones, with GRV 94 cross sections being about 5 to 10\% larger for $E_\nu
\gtrsim 10^9\ {\rm{GeV}}$. This is due to the fact that the GRV 94 sea and
gluon densities are slightly steeper at small-$x$ than the GRV 98 ones. Notice
that such a difference is within the typical uncertainty of about 20\% at
$x\simeq 10^{-8}$ due to factorization scale ambiguities \cite{15}. The same
holds true for the results in fig.\ 4 based on the CTEQ4-DIS densities as used
and extrapolated by GQRS 98 \cite{6} and the CTEQ3-DIS distributions \cite{24}
which provide equally reasonable and acceptable estimates of total neutrino
cross sections \cite{2}, except perhaps at highest energies where
GQRS 98(CTEQ4-DIS) possibly slightly underestimates them. The situation changes
if we turn to the results based on the remaining extrapolated parton
parametrizations in fig.\ 4: The FMR \cite{1} extrapolations and the
fixed-power extrapolated \cite{2} MRS-G distributions \cite{27} result in
noticeably larger cross sections, up to almost a factor of 2, at large
energies as compared to our dynamical QCD extrapolations and predictions. On
the other hand, the EHLQ-DLA distributions \cite{28}, extrapolated \cite{2} to
$x<10^{-4}$ using the 'double logarithmic approximation' (DLA), strongly
underestimate the cross sections almost throughout the whole high-energy range
shown in fig.\ 4.

These results indicate that UHE (anti)neutrino nucleon total cross sections can
be calculated with an uncertainty of about $\pm$20\% at highest neutrino
energies of $10^{12}\ {\rm{GeV}}$ which requires a reliable knowledge of
parton distributions at $x=10^{-8}$ to $10^{-9}$ and $Q^2=M_W^2$
\cite{15}. This is considerably more accurate than the estimated uncertainty
factor of $2^{\pm 1}$ derived from ad hoc extrapolations of fitted parton
distributions to very small-$x$ beyond the region of their validity
\cite{2,6}. Thus all recent estimates of cosmic UHE neutrino event rates
\cite{2,3,4,6} based on the dynamical GRV densities \cite{13,14,15} and the
properly extrapolated CTEQ distributions \cite{24,25}, or parton densities
with an effective similar small-$x$ behavior, appear to be realistic and
accurate to within 20\%; this is in contrast to estimates based on
extrapolated versions of FMR \cite{1} and MRS-G \cite{2,4} or
MRS-${\rm{D}}_-^\prime$ \cite{6} distributions, for example, which result in
too large rates at highest neutrino energies of about $10^{21}\
{\rm{eV}}$. These conclusions hold of course mainly for downward event rates,
since the upward muon event rates are obviously rather insensitive to the
particular choice of parton distributions due to the compensating attenuation
of neutrinos as they pass through the Earth \cite{2,6}.

\section*{Acknowledgements}
This work has been supported in part by the
'Bundesministerium f\"{u}r Bildung, Wissenschaft, Forschung und
Technologie', Bonn.   


\newpage
\section*{Figure Captions}
\begin{description}
\item[Fig.\ 1] 
The light quark sea and gluon distributions in NLO at $Q²=M_W^2$. The
dynamical GRV 94 \cite{14} and GRV 98  \cite{15} predictions have been
obtained from  evolving the respective valence-like input at $Q^2=Q_0^2$ which
is shown for illustration for the GRV 98 densities by the two curves at
$Q_0^2=0.40\ {\rm{GeV}}^2$. For comparison, the extrapolations of the CTEQ3
\cite{24} and CTEQ4 \cite{25}\footnotemark[3] distributions are shown as well.

\item[Fig.\ 2] 
Charged current $\nu N$ cross sections as calculated in NLO. In the case of
GRV 98 densities \cite{15}, heavy quark contributions are treated in a fully
massive way based on fixed-order perturbation theory as explained in the text,
whereas the GRV 92 distributions \cite{13} refer to a massless (resummed)
treatment of heavy quark flavors according to eq.\ (5). The GQRS 98 \cite{6}
calculation is based on the CTEQ4-DIS densities extrapolated to $x<10^{-5}$
using a fixed power in $x$ as given at $x=10^{-5}$ and $Q^2=M_W^2$; this
extrapolation method results in cross sections about 20 \% larger at
$E_\nu=10^{12}\ {\rm{GeV}}$ than the CTEQ4-DIS densities \cite{25}\footnotemark[3]
shown in fig.\ 1. The HERA measurement is taken from ref.\ \cite{26}.

\item[Fig.\ 3] 
The dependence of $\sigma_{CC}^{\nu N}$ on the small-$x$ region, i.e.\ on the
lower integration limit $x_{min}\leq x$ in eq.\ (1), using the NLO GRV 98
parton distributions \cite{15}.

\item[Fig.\ 4] 
Ratio of charged current cross sections using different parton distribution
functions (PDF) as compared to the GRV 98 distributions \cite{15}. The PDF
refer to GRV 94 \cite{14}, GQRS 98 (CTEQ4-DIS) \cite{6}, GQRS 96 (CTEQ3-DIS,
MRS-G) \cite{2}, FMR \cite{1} and EHLQ-DLA \cite{2}.

\end{description}

\newpage
\textheight 22.0cm
\textwidth 16cm
\oddsidemargin 0.0cm
\evensidemargin 0.0cm
\topmargin 0.0cm

\pagestyle{empty}
\begin{figure}
\vspace*{-1cm}
\hspace*{-1.5cm}
\epsfig{figure=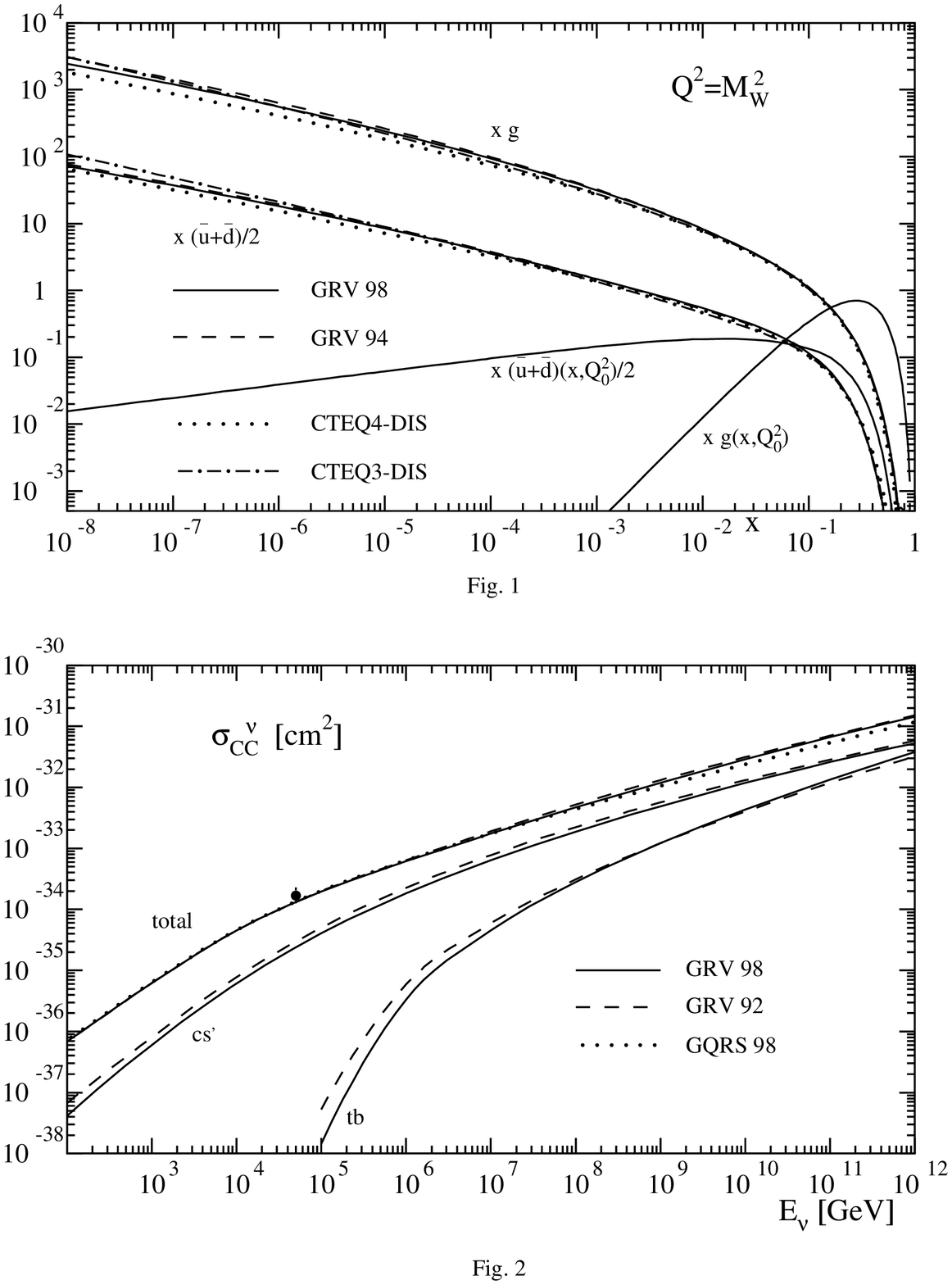,width=20cm}
\end{figure}
\newpage
\begin{figure}
\vspace*{-1.5cm}
\hspace*{-2.5cm}
\epsfig{figure=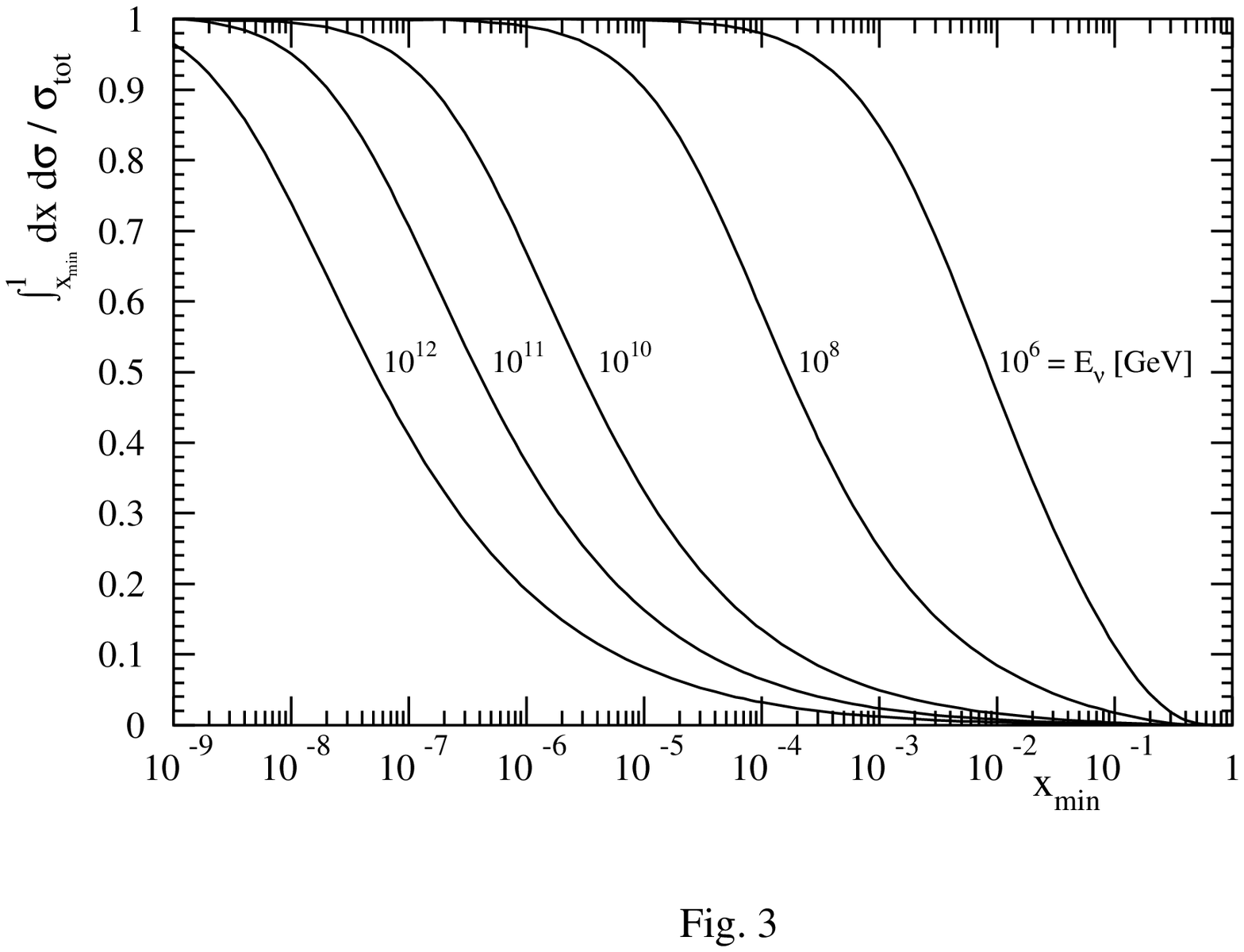,width=20cm}
\end{figure}
\newpage
\begin{figure}
\vspace*{-1.5cm}
\hspace*{-2.5cm}
\epsfig{figure=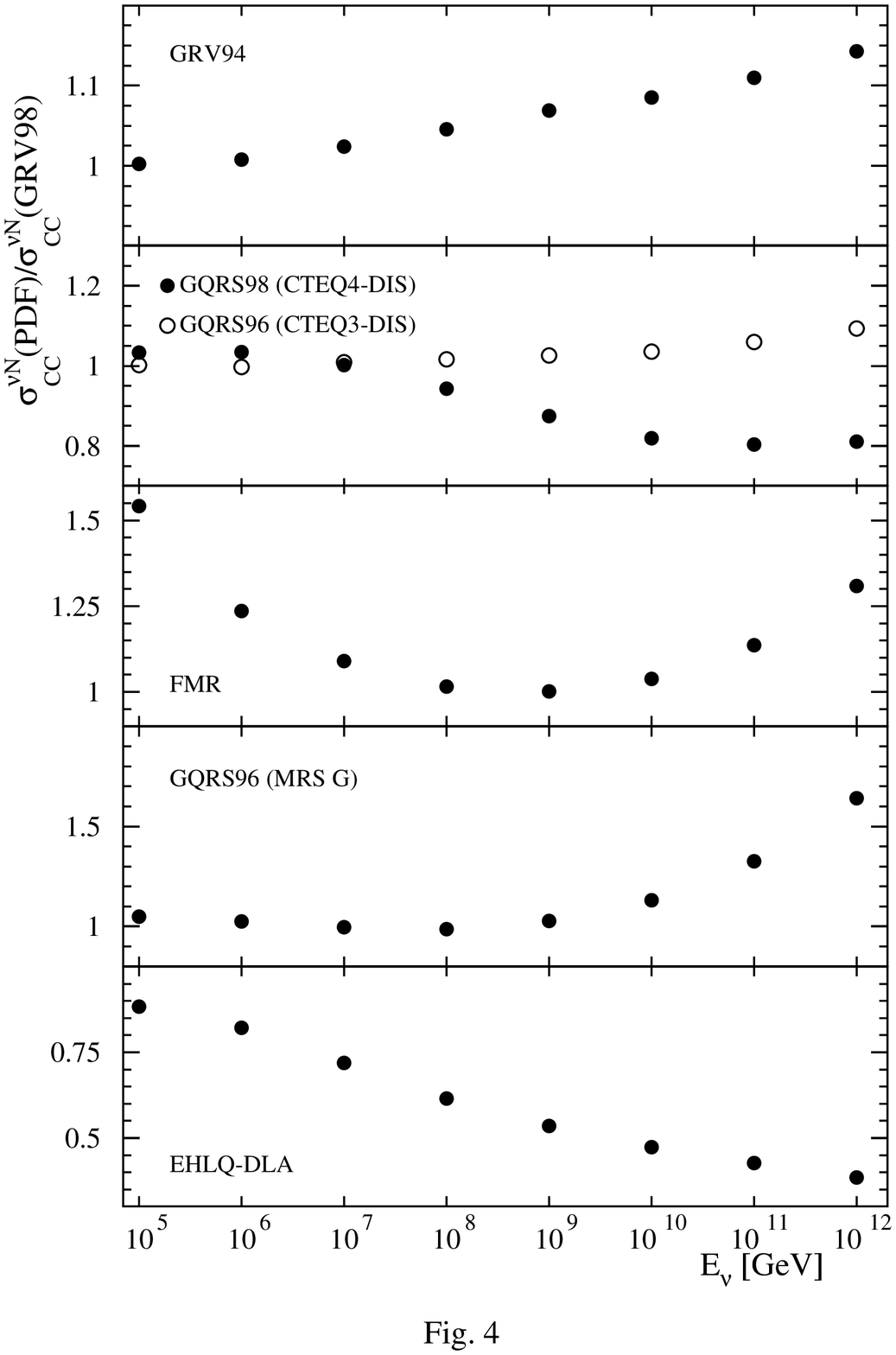,width=20cm}
\end{figure}
\end{document}